\documentclass[aps,prl,amsmath,amssymb,onecolumn,nofootinbib,showpacs,tightenlines,14pt]{revtex4}

\usepackage{epsfig,graphicx}
\usepackage{bm}
\newcommand{\be}{\begin{equation}}
\newcommand{\ee}{\end{equation}}
\newcommand{\bea}{\begin{eqnarray}}
\newcommand{\eea}{\end{eqnarray}}

\begin{document}
\title{Spherically symmetric massive scalar fields in GR}
\author{Mohammad Mehrpooya}
\affiliation{Department of Mathematics, Faculty of Science, Zabol
University, Zabol, Islamic Republic of Iran.}
\author{D.Momeni}
\email{dmomeni@phymail.ut.ac.ir} \affiliation{ Department of
Physics,Faculty of science,Islamic Azad University, Karaj
Branch.Iran, Karaj, Rajaei shahr, P.O.Box: 31485-313 }

\begin{abstract}
 First we review some of the attempts made to find exact spherically symmetric
solutions of Einstein field equations in the presence of scalar
fields .Wyman solution in both static and non static scalar field is
discussed briefly and it is show that why in the case of non static
homogenous matter field, static metric can not be represented in
terms of elementary functions. We mention here that if our spacetime
be static, according to EFE there is two option for choose scalar
field matter: static (time independent) and non static (time
dependent). All these solutions are limited to the minimally coupled
massless scalar fields and also in the absence of the cosmological
constant . Then we show that if we are interesting to have a
homogenous isotropic scalar field matter one can construct a series
solution in terms of scalar field's mass and cosmological constant.
This metric is static and posses a locally flat case as a special
chooses of mass of scalar field and can be interpreted as an
effective vacuum. Therefore mass of scalar field eliminates any
locally gravitational effect as tidal forces. Finally we describe
why this system is unstable in the language of dynamical systems.
\end{abstract} \maketitle
\section{\label{sec:level1}I.BACKGROUND}
Static spherically symmetric spacetimes are the simplest ones to
study. Bergmann and Leipnik [1] are the first who studied
solvability of field equations in the presence of scalar fields.
They were not able to find an explicit solution due to their choice
of a Schwarzschild like coordinate. Buchdahl in a series of papers
[2–4] introduced and developed the reciprocal metrics idea as a
generating method to construct new vacuum or non vacuum static
solutions as a simple transformation of a known vacuum or non vacuum
one. The higher dimensional extension of Buchdahl family has been
introduced by Xanthopolous and Zannias [5]. Later Wyman [6] attacked
this problem in a general framework and showed that apart from the
Buchdahl family of solutions there is another family of static
spherically symmetric solutions in which the scalar field is not
static. Later critical phenomena in gravitational collapse were
discovered by Choptuik [13–15] in the model of a spherically
symmetric, massless scalar field minimally coupled to general
relativity . The scalar field matter is both simple, and acts as a
toy model in spherical symmetry for the effects of gravitational
radiation. Important analytical studies of gravitational collapse in
this model have been carried out by Christodoulou [16–22].For other
numerical work on this model, see [23–27]. In this work after a
brief review of some of the previous studies we show that there is
no closed form of Wyman class and then in section III first we
derive EFE for a massive minimally coupled scalar field in the
presence of cosmological constant in a spherically symmetric space
time. Then by supposing that our scalar field has small time
variations such that we can disregard it's time dependency, and
corresponding to it a static metric, we derive a series solution for
our model. We mention here that if we assume that our field remains
static, pressure and energy density describe a stiff matter and our
spacetime posses homogeneity . This series solution has positive
spatial parity and by choosing a suitable mass of scalar field is
locally flat. Finally we study the stability of these solutions as a
dynamical system and as an important result we show that Wyman
minimally coupled solutions is unstable in the language of dynamical
systems.

\section{II.NON INTEGRABILITY OF NON STATIC SCALAR FIELDS IN SPHERICALLY
SYMMETRIC SPACETIMES}
  We start from  the Lagrangian of a static massless minimally
 coupled spherically symmetric scalar field with action:
\begin{eqnarray}
S=\int (-g)^{\frac{1}{2}}(R+\mu
g^{\mu\nu}\phi_{;\mu}\phi_{;\nu})d^{4}x
\end{eqnarray}
where in it  ($\mu,\nu=0,1,2,3$).
 We can write the static spherically symmetric solution
in schwarzschild gauge as:
\begin{eqnarray}
ds^{2}=-e^{\nu}dt^{2}+e^{\lambda}dr^{2}+r^{2}(d\vartheta^{2}+sin^{2}\vartheta
d\varphi^{2})
\end{eqnarray}
Einstein and scalar field equations are:
\begin{eqnarray}
\square \phi=0\\
R_{\mu\nu}-\frac{1}{2}R
g_{\mu\nu}=-\mu(\phi_{;\mu}\phi_{;\nu}-\frac{1}{2}g_{\mu\nu}\phi_{
;\alpha}\phi^{ ; \alpha})
\end{eqnarray}
 Buchdahl in series of papers   \cite{3,4} introduced  a new
 transformation which  we called  reciprocal
transformation. Imposing Buchdahl transformations to the
Schwarzschild  as  the seed metric ,we end up with  a one parameter
family of exact solution for a minimally coupled  massless static
scalar field .  As another example of the same  technique a static
solution with axial symmetry in weyl coordinates can be obtained
\cite{5} starting from the Weyl solution. The one parameter Buchdahl
solution in spherically symmetric coordinate is given by:
\begin{equation}\label{123}
\begin{aligned}
ds^{2}
  = -(1-\frac{2M}{r})^{\beta}dt^{2}+(1-\frac{2M}{r})^{-\beta}dr^{2}
     +r^{2}(d\vartheta^{2}+sin^{2}\vartheta d\varphi^{2})\\
\phi =\lambda \ln(1-\frac{2M}{r})\\
\beta =\pm \sqrt{1-2\mu \lambda^{2}}
\end{aligned}
\end{equation}
  Field equations are now simplified to;
\begin{eqnarray}
R_{\mu\nu}=-\mu \phi_{;\mu}\phi_{;\nu}
\end{eqnarray}
  Wyman  showed that  \cite{6} in Buchdahl formalism one family of solutions is
missing   and  Yilmaz   \cite{7},  Szekeres   \cite{8} and Buchdahl
solutions are all sub classes of his static family. Here we restrict
ourselves to Wyman's  non static spherically symmetric  scalar field
family of solution ; i.e the case in which
$\lambda=\lambda(r),\nu=\nu(r)$ , $\phi=\phi(t)$ and the metric
functions  (2) satisfy the following differential equation
equations:
\begin{eqnarray}
\begin{aligned}
\acute{\nu}+\acute{\lambda}=\mu r e^{\lambda-\nu}\\
\acute{\nu}-\acute{\lambda}=2(1-e^{\lambda})/r
\end{aligned}
\end{eqnarray}
Eliminating  $\nu$  we find the following for  $\lambda$;
 \begin{eqnarray}
\lambda^{''}+\frac{3\lambda^{'}}{r}(e^{\lambda}-1)+\frac{2}{r^{2}}(e^{\lambda}-1)(e^{\lambda}-2)=0
\end{eqnarray}
  Wyman has stated that the above equation (8) can not be integrated and by imposing
   suitable boundary conditions derived a
 series solution for the metric functions.  Now using results from
 differential equations we show that this family could not be
 written in closed form. Using the following change of variables :
\begin{eqnarray}
r=e^{t},u=\dot{\lambda}=\frac{d\lambda}{dt},x=e^{\lambda},
\end{eqnarray}
  Equation (8) is transformed into:
\begin{eqnarray}
u\frac{du}{dx}=u\frac{4-3x}{x}-\frac{2}{x}(x-1)(x-2)
\end{eqnarray}
  This differential equation is of  2-nd Abel type  (class A) \cite{9}. It is
  well known that  for a general equation of the  form:
\begin{eqnarray}
(y(x)+g(x))\frac{dy}{dx}=f_{2}(x)y(x)^{2}+f_{1}(x)y(x)+f_{0}(x)
\end{eqnarray}
  There are  solutions  only in the following  two special cases:\\
Case(1):  If\\
\begin{eqnarray}
\begin{aligned}
f_{0}(x)=f_{1}(x)g(x)-f_{2}(x)g(x)^{2}\\y(x)=-g(x)\\y(x)=e^{\int
f_{2}(x)dx}(c_{1}+\int(-f_{2}(x)g(x)+f_{1}(x))e^{-\int f_{2}(x)}dx)
\end{aligned}
\end{eqnarray}
Case(2):  If\\
\begin{eqnarray}
\begin{aligned}
f_{1}(x)=2f_{2}(x)g(x)-g(x)\acute{g(x)}\\
\alpha(x)=e^{-2\int
f_{2}(x)dx}\\\beta(x)^{2}=\alpha(x)^{2}g(x)^{2}+2\alpha(x)c_{1}+2\alpha(x)\int\frac{f_{0}(x)dx}{(\int
dx e^{\int f_{2}(x)dx})^{2}}\\
y(x)=-\frac{\alpha(x) g(x)+\sqrt{\beta(x)}}{\alpha(x)}
\end{aligned}
\end{eqnarray}
  Since  Wyman equation does not belong to any  of these cases
, one  can not derive an exact closed  form for it.

 In this section we  obtain a new series solution for a
massive-minimally coupled scalar field  in the presence of a
cosmological constant.According to Wyman we take our spherically
symmetric static spacetime metric  in Schwarzschild gauge,  and
require the metric  functions to  be regular near the  origin . So
that the scalar field and metric functions have a Taylor expansion
near the origin. The field equation are now given by:
\begin{eqnarray}
\begin{aligned}
R_{\mu\nu}-\frac{1}{2}Rg_{\mu\nu}-\Lambda g_{\mu\nu}=-\kappa
T_{\mu\nu}\\\kappa
T_{\mu\nu}=-\mu[\phi_{;\mu}\phi_{;\nu}-\frac{1}{2}g_{\mu\nu}(\phi_{,\alpha}\phi^{\alpha}-m^{2}\phi^{2})]
\end{aligned}
\end{eqnarray}
Where we have introduced a parameter  $\mu$  in  the energy-momentum
tensor to make sure one that our solution reduces to the  vacuum
solution in the absence of the scalar field. we take it 1 for
simplicity. The equivalent equation can be written as:
\begin{eqnarray}
R_{\mu\nu}=-\Lambda g_{\mu\nu}-\phi_{;\mu}\phi_{;\nu}+\frac{1}{2}
m^{2}\phi^{2} g_{\mu\nu}\end{eqnarray}
 Assuming  that our spherically symmetric
metric in the the Schwarzschild gauge is  of the general form:
\begin{eqnarray}
ds^{2}=e^{\nu}dt^{2}-e^{\lambda}dr^{2}-r^{2}(d\theta^{2}+sin^{2}
\theta d\varphi^{2})
\end{eqnarray}
 The  scalar field equation of motion  is:
\begin{eqnarray}
\Box\phi+m^{2}\phi=0,
\end{eqnarray}

  and equation (15) reduces to the following differential
  equations:
  \begin{eqnarray}
\begin{aligned}
-\frac{1}{4r}e^{-\lambda}(2r\nu^{''}+r(\nu^{'})^{2}-r\nu^{'}\lambda^{'}+4\nu^{'})=-\Lambda +\frac{1}{2} m^{2}\phi^{2}\\
\frac{1}{4r}(2r\nu^{''}+r(\nu^{'})^{2}-r\nu^{'}\lambda^{'}-4\lambda^{'})=
\Lambda e^{\lambda}- \acute{\phi}^{2}- \frac{1}{2} m^{2}\phi^{2}e^{\lambda}\\
 \frac{1}{2}e^{-\lambda}(-r\lambda^{'}-2e^{\lambda}+r\nu^{'}+2)= -\Lambda r^{2}- \frac{1}{2}
 m^{2}\phi^{2}r^{2}\\
 \frac{d^{2}\phi}{dr^{2}}+2(\frac{1}{r}+\frac{\nu^{'}-\lambda^{'}}{2})\frac{d\phi}{dr}-m^{2}e^{\lambda}\phi=0
 \end{aligned}
\end{eqnarray}
From  (18) after some little algebra we find:
\begin{eqnarray}
\begin{aligned}
\acute{\nu}+\acute{\lambda}=- r \acute{\phi}^{2}\\
\acute{\nu}-\acute{\lambda}=\frac{2}{r}[e^{\lambda}(1-\Lambda
r^{2})-1-\frac{1}{2} m^{2}\phi^{2}r^{2}e^{\lambda}]
\end{aligned}
\end{eqnarray}
 There is no simple reduction of this system to a one variable
solvable differential equation. Actually field  equation can be
writhen in the following complicated non linear form:
\begin{eqnarray}
\begin{aligned}
\acute{\phi}^{2}=\frac{h_{3}{\phi}^{'''}+3h_{2}{\phi}^{''}}{h_{0}},
\end{aligned}
\end{eqnarray}
  in which:
\begin{eqnarray}
\begin{aligned}
h_{0}=r{\phi}^{''}(r m^{2}\phi+{\phi}^{'}(-2+2\Lambda
r^{2}+m^{2}r^{2}\phi^{2}))\\
h_{3}=(-2+2\Lambda
r^{2}+m^{2}r^{2}\phi^{2}){\phi}^{''}+\frac{4}{3}m^{2}
r^{2}\phi{\phi}^{'}+\frac{4}{3}r{\phi}^{'}(m^{2}\phi^{2}+\frac{m^{2}}{2}+2\Lambda)+\frac{2}{3}m^{2}\phi\\
h_{2}=(4-4\Lambda r^{2}-2m^{2}r^{2}\phi^{2}){\phi}^{'}-2 m^{2} r\phi
\end{aligned}
\end{eqnarray}
    It does not seem one could be able  solve (20) analytically, so we try a semi-analytic solution
    by imposing the same  simple
boundary conditions on metric as Wyman did \cite{6}:
\begin{eqnarray}
\nu(0)=0,\lambda(0)=0,\phi(0)=q,\acute{\phi}(0)=0
\end{eqnarray}
 These  boundary conditions are chosen such that  we have a regular
solution at the  origin.  q is interpreted as a new parameter   of
our two parameter $(q,m)$ family of exact solutions, apart from the
mass of the scalar field and the cosmological constant. Applying the
above boundary conditions in equation (19) and  taking successive
derivatives of equations and evaluating  them at the origin we can
write the following series for the scalar field and the  metric
functions \cite{12}:
\begin{eqnarray}
\begin{aligned}
\phi(r)=q(1+\sum_{n=2,4,6,...}a_{n}(q^{2})r^{n}
)\\e^{\nu}=1-\sum_{n=2,4,6,...}b_{n}(q^{2})r^{n}
\\e^{\lambda}=1+\sum_{n=2,4,6,...}c_{n}(q^{2})r^{n}
\end{aligned}
\end{eqnarray}
 We note that only even values of  n contribute in the  solution.
 This means that our functions remain unchanged under parity transformation  $r\rightarrow -r$. We can
 conclude that our metric functions can be written  in general form
  $X(r)=1\pm f(m^{2}q^{2},\Lambda r^{2})$.
  The first few values of the coefficients in (23) are  give  below:
\begin{eqnarray}
\begin{aligned}
a_{2}(x)=\frac{1}{6} m^{2}\\a_{4}(x)=\frac{1}{4!}(\frac{1}{3}
m^{4}x+\frac{1}{5} m^{4}+\frac{2}{3} \Lambda
m^{2})\\b_{2}(x)=\frac{1}{6}[ m^{2}
x+2\Lambda]\\b_{4}(x)=\frac{1}{60}[ m^{4} x]\\c_{2}(x)=\frac{1}{6}[
m^{2} x+2\Lambda]\\c_{4}(x)=\frac{1}{24}[\frac{16}{15} m^{2}
x+\frac{2}{3} m^{4} x^{2}+\frac{8}{3} m^{2}\Lambda
x+\frac{8}{3}\Lambda^{2}]
 \end{aligned}
\end{eqnarray}
  If  $q=0$ the solution reduces to  that of the series expansion for  de-Sitter solution . It is clear that there is no
consistent Wyman  solution of 2-nd class. Calculating all the Rieman
Tensor components  we observe that the only non zero  component up
to 4-th order in r  (near the origin) is given by:
\begin{eqnarray}
R^{0}_{101}=\frac{1}{6}\mu m^{2}q^{2}-\frac{1}{3}\Lambda
\end{eqnarray}
 So if the mass of the scalar field satisfies
$m=q^{-1}\sqrt{2\Lambda}$ this component vanishes and the spacetime
becomes  flat and every  local effect of gravity as tidal forces
vanishes.
\section{massive scalar field as a Dynamical system}
In this section we apply a  powerful method to analyze the stability
of the scalar field solutions obtained in the previous section .
This method for  an autonomous ODE  has already been employed by
Coley    \cite{10} in the context of cosmology. To do so we
introduce some concepts from nonlinear complex systems   \cite{11}.
In the language of  dynamical systems our differential equation
could be represented as follows :
\begin{eqnarray}
\begin{aligned}
\dot{x}=f(t,x)\\ f:[0,\infty)\times D\longrightarrow
\mathbb{R}^{n}\\ D=\{x \in\mathbb{R}^{n}   \mid      \parallel
x\parallel_{2}<0\}
 \end{aligned}
\end{eqnarray}
 We also assume that $x=0$ be the  equilibrium point of the system at $t=0$ that is:
\begin{eqnarray}
f(t,0)=0,\forall t\geq 0
\end{eqnarray}
 We assume  that the Jacobian matrix ; $[\frac{\partial f}{\partial
x}]$   is bounded  with respect to t on the set D and  smoothly
satisfies Lipschit'z lemma ,  that is:
\begin{eqnarray}
\parallel f(t,x)-f(t,y) \parallel \leq L \parallel x-y
\parallel,\parallel x \parallel _{p}=(\sum^{p}_{i}\mid x_{i} \mid
^{p})^{\frac{1}{p}},1\leq p <\infty
\end{eqnarray}
 One  can state  the  following   theorems about this system.\\
 Theorem   (1): In the neighborhood of any stability point the  system can be approximated by a linear one i.e;:
\begin{eqnarray}
\dot{x}=A(t) x
\end{eqnarray}
  Theorem  (2): Assume that $x=0$ be an equilibrium point of the  nonlinear
 system    $\dot{x}=f(t,x)$    that is :
\begin{eqnarray}
\begin{aligned}
  f:[0,\infty)\times D\longrightarrow
\mathbb{R}^{n}\\ D=\{x \in\mathbb{R}^{n} \mid   \parallel
x\parallel_{2}<r\},
 \end{aligned}
\end{eqnarray}
  and define the  matrix;   $ A(t)=\frac{\partial f(t,x)}{\partial x}|_{x=0}
$  . If the origin is an  exponential stable equilibrium point of
the linear system    (29)     then this point is the exponential
stable equilibrium point of the  non-linear system (26).\\
 Now we apply these theorems to our system. First we rewrite field
equations through the following variables:,
\begin{eqnarray}
x_{1}=\nu(r),x_{2}=\lambda(r),x_{3}=\phi(r),x_{4}=\frac{d\phi(r)}{dr},
\end{eqnarray}
  as follows
\begin{eqnarray}
\begin{aligned}
\acute{x_{1}}=\frac{e^{x_{2}}-1}{r}+\Lambda r e^{x_{2}}-\frac{1}{2}
r(x_{4}^{2}+m^{2}
e^{x_{2}}x_{3}^{2})\\\acute{x_{2}}=-\frac{e^{x_{2}}-1}{r}-\Lambda r
e^{x_{2}}-\frac{1}{2} r(x_{4}^{2}-m^{2}
e^{x_{2}}x_{3}^{2})\\\acute{x_{3}}=x_{4}\\\acute{x_{4}}=[\frac{-e^{x_{2}}+1}{r}-\Lambda
r e^{x_{2}}+\frac{1}{2} m^{2} r
x_{3}^{2}e^{x_{2}}]x_{4}-m^{2}x_{3}e^{x_{2}}
 \end{aligned}
\end{eqnarray}
 Equilibrium point of the above   system lies at:
\begin{eqnarray}
x_{1}=x_{1},x_{2}=-\ln(1+\Lambda r^{2}),x_{3}=0,x_{4}=0
\end{eqnarray}
 This point of  4-dimensional phase space corresponds to
a   $t=cte$   hypersurface with constant negative intrinsic
curvature $k=-\Lambda$ with metric:
\begin{eqnarray}
ds^{2}_{t=cte}=\frac{dr^{2}}{1+\Lambda
r^{2}}+r^2(d\theta^{2}+sin^{2} \theta d\varphi^{2})
\end{eqnarray}
 We assume that the equilibrium  point   (solution)   is perturbed as
 $x_{i}=x_{i}^{0}+u_{i}$. The Jacobian matrix can be easily
calculated  and the result is:
\begin{eqnarray}
A_{ij}=[{\frac {\partial f_{i} }{\partial x_{j}}} \left( t,x \right)
]= \left[
\begin {array}{cccc} 0&{r}^{-1}&0&0\\\noalign{\medskip}0&{r}^{-1}&0&0
\\\noalign{\medskip}0&0&0&1\\\noalign{\medskip}0&0&{\frac {{m}^{2}}{1+
\Lambda\,{r}^{2}}}&0\end {array} \right]
\end{eqnarray}
  We can  find an exact solution for perturbations of our dynamical system
variables  (29)  according to the following system of ordinary
differential equations, using  (29) ;
\begin{eqnarray}
\begin{aligned}
\acute{u_{1}}=\frac{1}{r}u_{2}\\
\acute{u_{2}}=\frac{1}{r}u_{2}\\
\acute{u_{3}}=u_{4}\\
\acute{u_{4}}=\frac{m^{2}}{1+\Lambda r^{2}}u_{3}
\end{aligned}
\end{eqnarray}
 From these set of equations  we immediately  obtain:
\begin{eqnarray}
\begin{aligned}
\hspace{-50mm}u_{1}=c_{1}r+c_{2}\\
\hspace{-50mm} u_{2}=c_{1}r\\ u_{3}=(1+\Lambda
r^{2})(c_{1}F([a_{1},a_{2}],1/2,-\Lambda r^{2})+\\\hspace{-50mm}c_{2}F([a_{1}+1/2,a_{2}+1/2],1/2,-\Lambda r^{2})) \\
\hspace{-50mm}u_{4}=\frac{d u_{4}}{dr}
\end{aligned}
\end{eqnarray}
 In  which:
\begin{eqnarray}
\begin{aligned}
 F(n, d, z) =\sum^{\infty}_{k=0}\frac{z^k}{k!} \prod^{p}_{i=1}(n_{i})_{k}
 \prod^{q}_{j=1}(d_{j})_{k}\\
 (d_{j})_{k}=\frac{\Gamma(k+d_{j})}{\Gamma(k)}\\
 n = [n_{1}, n_{2}, ...], p = nops(n), d = [d_{1}, d_{1}, ...],q = nops(d)\\
     a_{1}=1/4\,{\frac {3\,\sqrt {\Lambda}+i\sqrt
{4\,{m}^{2}-\Lambda}}{\sqrt { \Lambda}}}\\
a_{2}=-1/4\,{\frac {-3\,\sqrt {\Lambda}+i\sqrt
{4\,{m}^{2}-\Lambda}}{\sqrt { \Lambda}}},
\end{aligned}
\end{eqnarray}
 and the symbol $nops$ means the number of operands of an
 expression. If some  $n_{i}$  is a non-positive integer,
 the series is finite (that is, F(n, d, z) is a polynomial in z).
If some  $d_{j}$  is a non-positive integer, the function is
undefined for all non-zero z, unless there is also a negative upper
parameter of smaller absolute value, in which case the previous rule
applies. For the remainder of this description, assume no  $n_{i}$
or  $d_{j}$  is a non-positive integer. When $p \leq q$ , this
series converges for all complex z, and hence defines F(n, d, z)
everywhere. When $p = q+1$ , the series converges for  $|z| < 1$ .
  $F(n, d, z)$   is then defined for $ |z|\geq 1$
 by analytic continuation. The point $z=1$  is a branch point, and
the interval  $(1,\infty)$  is the branch cut. When  $p > q+1$  the
series diverges for all  $z\neq 0$ . In this case, the series is
interpreted as the asymptotic expansion of   $F(n, d, z)$  around $z
= 0$ . The positive real axis is the branch cut.\\
  Now the main result comes from the fact that the solution  (37) is not asymptotically convergent  i.e;
\begin{eqnarray}
u_{i}(r\rightarrow \infty)=\infty
\end{eqnarray}
  So although  we treated the system in the linear approximation, we deduce  that it is not stable . Therefor
  it is guaranteed  that our nonlinear system is not stable too. For a massless
(Wyman) solution in the absence of  the cosmological constant  we
can refer to the  system of equations (36) and its solutions  (37).
 Obviously  in this case  the  perturbations diverge and so this family is not
stable as well.
\section{Acknowledgement}
D. Momeni thanks Anzhong Wang for useful comments and valuable
suggestions. Finally, the editor of IJMPA , Chee-Hok Lim and the
anonymous referees made excellent observations and suggestions which
resulted in substantial improvements of the presentation and the
results. The authors would like to thank the Islamic Azad
University, Karaj branch for kind hospitality and support. This work
is supported by Islamic Azad University (Karaj branch). Address and
affiliation at the time this paper was written: Department of
Physics, Islamic Azad University, Karaj branch, 31485-313, Iran


\begin{thebibliography}{35}
\bibitem{1}
Q. Bergmann and R. Leipnik. Phys. Rev. 107. 1157 (1957)
\bibitem{2}
H. A. Buchdahl, Quart. J. Math. (Oxford) 5, 116 (1954)
\bibitem{3}
H. A. Buchdahl, Australian J. Phys. 9, 13 (1956)
\bibitem{4}
H. A. Buchdahl. Phys. Rev. 115, 1325-1328 (1959)
\bibitem{5}
B. C. Xanthopolous and T. Zannias. Phys. Rev. D. vol 40, Num 8
(1989)
\bibitem{6}
M. Wyman. Phys. Rev. D. vol 40, Num  8 (1981)
\bibitem{7}
H. Yilmaz. Phys. Rev. 111. 1417 (1958)
\bibitem{8}
G. Szekeres. Phys. Rev. vol  97, Num 1 (1955)
\bibitem{9}
  E. Kamke, Handbook of Ordinary Differential Equations [Russian
  translation], Vol. 1, Nauka, Moscow (1965).\\
  I.G. Petrovskii,Lectures on the Theory of Differential Equations [in Russian],
  Nauka, Moscow (1970)\\
  E . Kamke. Differentialgleichungen: Losungsmethoden und Losungen. New York: Chelsea Publishing Co,
  1959. p.26\\
 D. Zwillinger. Handbook of Differential Equations, 2nd edition. Academic Press,
 1992.\\
 E. S. Cheb-Terrab, L .G . S. Duarte and L . A. C. P. da Mota.
 Computer Physics Communications 101 (1997): 254
\bibitem{10}
A. A. Coley. arxiv: gr-qc /9910074. v1. 21 Oct 1999
\bibitem{11}
    E.N. Rozenvasser, Nonlinear Oscillations [in Russian], Nauka, Moscow
 (1969)\\
 S. Wiggins, Introduction to Applied Non linear Dynamical Systems and
chaos  (Springer 1990).
\bibitem{12}

 We take limit from both sides of equations(19) and using
boundary conditions mentioned in (22) we deduced that  $
\acute{\nu}(0)=\acute{\lambda}(0)=0 $  . Also from equation (18) we
immediately obtain  $ \phi^{''}(0)=\frac{1}{3}m^{2}q $  .
Differentiation from equations(19) with respect to r and taking
limit $r\rightarrow 0$   we obtain $\nu^{''}(0)=\frac{-1}{3}\mu
m^{2} q^{2} +\frac{2}{3}\Lambda, \lambda^{''}(0)=\frac{1}{3}\mu
m^{2} q^{2} -\frac{2}{3}\Lambda, \phi^{'''}(0)=0 $    .We seek that
for all functions of our model $X^{2n+1}(0)=0,n\geq 0 $ .
Unfortunately since there is no simple one variable differential
equation which from it,we can obtain a recursion relation this
assumption can not be proved analytically.
\end{thebibliography}
\end{document}